\documentclass[10pt]{article}

\usepackage{graphicx}
\usepackage{algorithm}
\usepackage{algpseudocode}
\usepackage{url}
\usepackage{verbatim}
\usepackage{listings}
\usepackage{color}
\usepackage{tabularx}
\usepackage{float}
\usepackage{caption}
\usepackage{subcaption}
\usepackage{mathptm}
\usepackage{booktabs}
\usepackage{multirow}

\title{ImageCL: An Image Processing Language for Performance Portability on Heterogeneous Systems\footnote{This is a pre-print version of an article to be published in the Proceedings of the 2016 Conference on High Performance Computing and Simulation. For personal use only.}}

\author{Thomas L. Falch and Anne C. Elster \\
Department of Computer and Information Science,\\ Norwegian University of Science and Technology \\
Trondheim, Norway \\
and Institute for Computational Engineering and Science, \\
University of Texas at Austin, USA\\
Email: thomafal@idi.ntnu.no, elster@ntnu.no
}
\date{}

\begin{document}

\lstset{basicstyle=\small\ttfamily,captionpos=b,language=C,morekeywords={complex,Image,idx,idy}}
\providecommand{\keywords}[1]{\small{\textbf{Keywords: }} #1}

\maketitle

\begin{abstract}
    Modern computer systems typically conbine multicore CPUs with accelerators like GPUs for
    inproved performance and energy efficiency. However, these systems suffer from poor performance
    portability -- code tuned for one device must be retuned to achieve high performance on another.
    Image processing is increasing in importance , with applications 
    ranging from seismology and medicine to Photoshop.

    Based on our experience with medical image processing, we propose ImageCL, a high-level domain-specific
    language and source-to-source compiler, targeting heterogeneous hardware. ImageCL
    resembles OpenCL, but abstracts away performance optimization details, allowing the programmer
    to focus on algorithm development, rather than performance tuning. The latter is left to our
    source-to-source compiler and auto-tuner. From high-level ImageCL kernels, our source-to-source
    compiler can generate multiple OpenCL implementations with different optimizations applied. We
    rely on auto-tuning rather than machine models or expert programmer knowledge to determine which
    optimizations to apply, making our tuning procedure highly robust.  Furthermore, we can generate
    high performing implementations for different devices from a single source code, thereby
    improving performance portability. 
    
    We evaluate our approach on three image processing benchmarks, on different GPU and CPU devices, and
    are able to outperform other state of the art solutions in several cases, achieving speedups of
    up to 4.57x.
\end{abstract}


\section{Introduction}
\label{intro}
Heterogeneous computing, where devices with different architectures and performance
characteristics, such as CPUs, GPUs and accelerators are combined, are growing in popularity.
These systems promise increased performance combined with reductions in power consumption, but 
the increasing complexity and diversity results in issues with programmability, such as poor \emph{performance
portability}.

Standards like OpenCL offer functional portability for heterogeneous systems.
However, performance portability, making code achieve good performance when
executed unaltered on different devices, remains problematic. This is
particularly true for highly different devices, such as a CPU and a GPU, but
also for two GPUs from the same vendor \cite{FALCH2}.

Auto-tuning can potentially improve the situation. In this setting, auto-tuning involves
automatically evaluating different candidate implementations and selecting the best one for a given
device. Thus, empirical data is used to find the best code, rather than relying on potentially
faulty and incomplete programmer intuition or compiler machine models. Performance portability is
improved, since porting to a new device simply requires the auto-tuning to be re-done. While a
high number of candidate implementations can lead to prohibitive search times, analytical
performance models, or techniques based on machine learning, as proposed in our previous
work \cite{FALCH2}, can be used to speed up the search.

For this kind of auto-tuning to be truly automatic, it must include the generation of candidate
implementations. Writing these manually is tedious, time consuming, and error prone. In some cases,
simply changing the value of performance critical variables, such as the work-group size in OpenCL,
is enough to generate what effectively becomes different implementations.  However, many important
optimizations require more substantial code changes. For instance, whether to use image or local
memory in OpenCL effectively requires two versions of the source code. We propose using
high level languages and source-to-source compilers to automatically generate candidate
implementations with potentially widely different source code, thus enabling an end-to-end
auto-tuning approach to overcome heterogeneous performance portability.

In this paper, we present the ImageCL language and its source-to-source compiler.  ImageCL programs
resemble simplified versions of OpenCL kernels. Algorithms written in ImageCL are analysed by
our compiler, and different optimizations are applied to generate a large number of different
candidate implementations, covering an extensive optimization space. We then use the machine
learning based auto-tuner from our previous work \cite{FALCH2} to pick a good implementation for a
given device.  Thus, an algorithm can be written once, and our auto-tuning compiler tool-chain can
be used to generate high performing implementations for any device supporting OpenCL, thus improving
performance portability.

Image processing is an increasingly important domain, with applications ranging from medicine and
seismology to Photoshop.  ImageCL has been designed to work with the FAST \cite{SMISTAD2} framework.
FAST allows the user to connect filters to form image processing pipelines, which can be executed on
heterogeneous systems. ImageCL can be used to write a single such filter, which can be retuned to
achieve high performance if scheduled on different devices. While ImageCL can be considered a domain
specific language for image processing, it retains the generality of OpenCL.

This paper is structured as follows: The next section provides background information, while
Section~\ref{related} reviews related work.  Section~\ref{compiler} presents a high level overview
of how our approach can be used to achieve performance portability. In Section~\ref{language}, the
ImageCL language and the implementation of the source-to-source compiler is described. Results are
presented in Section~\ref{results}, and discussed in Section~\ref{discussion}. Finally,
Section~\ref{conclusion} concludes and outlines possible future work.

\section{Background}
\label{background}

In this section, we provide background information on heterogeneous computing, OpenCL, and the FAST
framework, which our language and has been designed to work with.

\subsection{GPU Computing and OpenCL}
As several of our optimizations target GPU specific
features, we will review their architecture here. For details, see e.g. \cite{SMISTAD}.

Modern GPUs are built up of large numbers of processing elements, which are combined into compute units
\footnote{Here we use OpenCL terminology. Nvidia uses the terms CUDA cores and
streaming multiprocessors, AMD stream processors and compute units.}.  The processing
elements of a compute unit work in a SIMD fashion, executing instructions in lock step. Discrete
GPUs have large, relatively slow, DRAM memories (separate from the system's main memory) known as
global memory. On newer GPUs, global memory is often cached. In addition, they have fast,
on-chip, scratch-pad memory, which can be used as a user managed cache.  Furthermore, they have
texture memory, which is cached, and optimized for access patterns with 2D and 3D spatial locality,
as well as constant memory which is read-only and designed for high performance when accessed by many
threads concurrently.
 
OpenCL is emerging as a standard for heterogeneous computing, and is supported by many
major hardware vendors. OpenCL code is divided into host code and kernels. The host
code sets up and launches kernels on a device like a GPU or the
same CPU the host code is running on. Kernels are executed in parallel by multiple threads known
as work-items, which are organized into work-groups. On a GPU, work-groups are mapped to 
compute units, while work-items are mapped to processing elements, on a CPU they are
mapped to the CPU cores.  OpenCL has several logical memory spaces: local memory (mapped to the
fast on-chip memory on GPUs), image memory (mapped to the GPU texture memory), and constant memory
(mapped to the hardware constant memory on GPUs). On the CPU, these memory spaces are all
typically mapped to main memory.

\subsection{Image Processing and FAST}
\label{fast}
FAST \cite{SMISTAD2} is a recent framework that allows the user to create image processing
applications by connecting together pre-implemented filters to form a pipeline. Each filter take one
or more images as input, and produce one or more images as output. The filters are written in OpenCL
for GPUs or C++ for CPUs, and can in principle provide multiple implementations for different
devices. If executed on a system with multiple devices such as GPUs and CPUs, each filter in the
pipeline can be scheduled to run on any of the available devices, with memory transfers handled
automatically, thus taking full advantage of the heterogeneous system to achieve good performance.

FAST makes it easy to write heterogeneous image processing applications from existing filters, but
writing these filters is challenging due to performance portability. Each filter may be executed on
different devices depending upon the machine it is executed on and the pipeline it is a part of, and
must therefore often provide multiple different implementations tuned for different devices to ensure
optimal performance on all of them.

\section{Related work}
\label{related}

Auto-tuning is an established technique, used successfully in high performance libraries
like FFTW \cite{FFTW} for FFTs and ATLAS \cite{ATLAS} for linear algebra, as well as for
bit-reversal \cite{ELSTER}. Methods to reduce the search effort of auto-tuning, such as analytical
models \cite{YOTOV}, or machine learning \cite{BERGSTRA}, have been developed. 

Poor OpenCL performance portability has been the subject of many works. Zhang et al. \cite{ZHANG}
identified important tuning parameters greatly affecing performance.  Pennycook et al.
\cite{PENNYCOOK} attempted to find application settings that would achieve good performance accross
different devices. Auto-tuning approaches have also been proposed in \cite{FALCH2,NUGTEREN}, but
required the OpenCL code to be manually parameterized.

Directive based approaches, including OpenMP 4.0 and OpenACC takes the
level of abstraction even higher, and allow users to annotate C code with directives to offload
parts to accelerators or GPUs.  In contrast, our work is focused on making it simpler to write the
code for a single kernel, using a implicitly data parallel language, rather than offloading and
parallelizing serial CPU code.  Furthermore, our approach is better suited for integration with
frameworks such as FAST, which only requires the kernels to be written.

High-level and domain-specific languages (DSL) have a long history \cite{DEURSEN}.  While primarily
designed to ease programming, but the domain specific knowledge can also be used to generate
optimized code. For example, the Delite \cite{SUJEETH} framework has been used to develop
performance oriented DSLs for multi-core CPUs and GPUs, such as OptiML for machine learning, and
OptiGraph for graph processing.

High performance DSLs for image processing have also been proposed, and many of these works resemble
our own. Halide \cite{RAGAN2013} is a DSL embedded in C++, particularly targeting graphs of stencil
operations for image processing. Halide separates the algorithm, specified in a purely functional
manner, from the \emph{schedule} which specifies how the calculations should be carried out,
including tiling, parallelization, and vectorization.  Optimization is done by changing the
schedule, without modifying the algorithm, or its correctness. Schedules can be hand-tuend, or
auto-tuned using stochastic search. GPUs can be targeted, but important GPU optimizations, such as
using specific memories, are hard or impossible to express.

HIPACC \cite{MEMBARTH2016} is another DSL for image processing, also embedded in C++, but with a more
traditional, imperative approach than Halide, and a larger focus on single filters rather than
pipelines. A source-to-source compiler can generate code for different back-ends, including OpenCL,
CUDA and C++. Domain specific knowledge, as well as data gathered from analysing the input algorithm
is combined with a architecture model to generate optimized code.  Optimizations that can be applied
include memory layout, use of the memory hierarchy, thread coarsening, and efficient handling of
boundary conditions. A heuristic is used to determine work-group sizes.

PolyMage \cite{MULLAPUDI}, like Halide, focuses on complete image processing pipelines, with a
functional style. It uses a model driven, optimizing compiler that only targets multi-core CPUs.

There is also a body of work on transforming naive , simplified CUDA or OpenCL kernels and into
optimized kernels using optimizations related to e.g. the memory hierarchy, memory coalescing, data
sharing and thread coarsening.  \cite{UENG, LIN, YANG}. While bearing semblance to our work, none of
these have features specifically suited for image processing, combine all the optimizations we
apply, or rely on auto-tuning.

Combining code generation or source-to-source compilers with auto-tuners has also been explored.
Khan et al. \cite{KHAN} used a script based auto-tuning compiler to translate serial C loop nests to
CUDA.  Du et al. \cite{DU} proposed combining a code generator for linear algebra kernels with a
auto-tuner to achieve performance portability. The PATUS \cite{CHRISTEN} framework can generate and
auto-tune code for stencil computations for heterogeneous hardware, using a separate specification
for the stencil and the computation strategy, similar to Halide.  It lacks the general purpose
capabilities of our work, and does not support all our optimizations.

\section{ImageCL and Auto-Tuning}
\label{compiler}

\begin{figure}[htbp]
    \centering
    \includegraphics[width=0.45\textwidth]{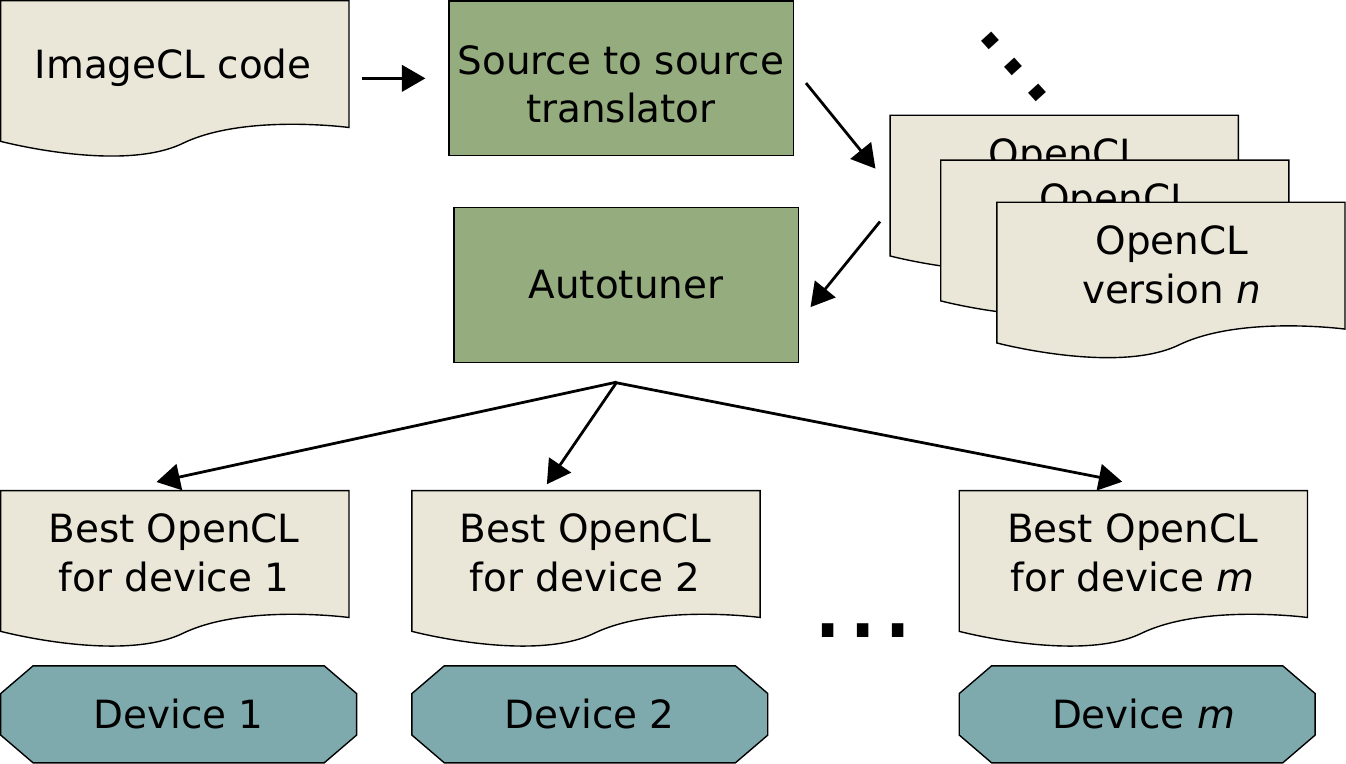}
    \caption{Overview of how the ImageCL source-to-source compiler would work
    with an auto-tuner.}
    \label{overview}
\end{figure}

 Starting with a description of the algorithm in ImageCL, our source-to
source compiler generates multiple candidate implementations in OpenCL, each with a different set of
optimizations applied. The auto-tuner then picks the best implementation for a given device. One can
thus easily generate multiple, high-performing versions for different devices from a single source,
achieving greater performance portability. This is illustrated in Figure~\ref{overview}

\begin{figure}[htbp]
    \centering
    \includegraphics[width=0.45\textwidth]{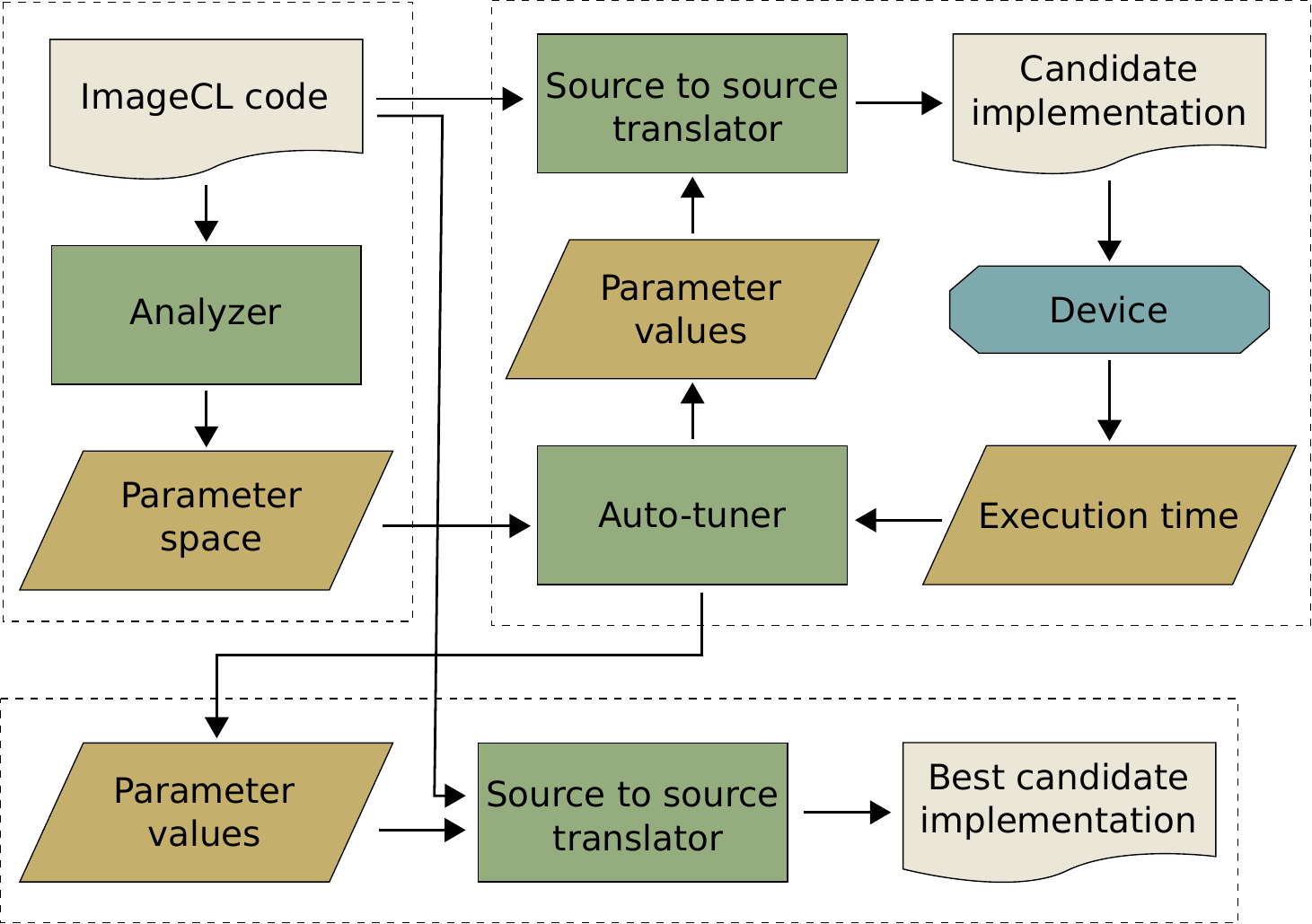}
    \caption{Detailed view of source-to-source compiler - auto-tuner interface .}
    \label{overview_detail}
\end{figure}

A more detailed view of how the source-to-source compiler works together with the auto-tuner to find
the best implementation for a given device can be found in Figure~\ref{overview_detail}.  Initially,
the ImageCL code is analyzed to find the potential optimizations, that is, the tuning parameters and
their possible values. Next, the auto-tuner explores the parameter space by selecting particular
values for the tuning parameters, generating the corresponding OpenCL code with the source-to-source
compiler, compiling it with the device compiler, and executing and timing it on the relevant device.
The procedure is repeated using some arbitrary search method until the auto-tuner arrives at what it
believes to be the best parameter values. Finally, the source-to-source compiler uses these
parameter values to generate the final implementation.

An analytical performance model or expert programmer knowledge could be used to determine parameter
values directly. However, we believe the complexity and rapid development of the hardware makes
auto-tuning a more robust option, and therefore intend our source-to-source compiler to be used with
an auto-tuner. While any general purpose auto-tuning framework can be used, our compiler was
designed with the auto-tuner of our previous work \cite{FALCH2} in mind.

Our auto-tuner uses a machine learning based performance model to to find promising parameter
configurations, to speed up the search. In particular, the auto-tuner will execute the code of
several randomly selected parameter configurations and record the execution times. This data is then
used to build an artificial neural network performance model, which can predict the execution time
of unseen configurations. The model is then used to predict the execution time of all possible
configurations, which can be done, even for large search spaces, since model evaluation is cheap. In
a second step, some of the configurations with the best predicted execution times are executed, and
the configuration with the best actual execution time of these is returned by the auto-tuner.

We generate OpenCL since it is supported by multiple vendors and targets a broad range of devices.
Compared to generating device specific assembly, it reduces the engineering effort, and allows us to
leverage optimizations performed by the device OpenCL compilers.

\section{The ImageCL Language}
\label{language}
The ImageCL programming language is designed to make it easy to write image processing kernels for
heterogeneous hardware, and be used together with the FAST framework. An example of the language is
shown in Listing~\ref{blur}, which implements a simple 3x3 box filter for blurring. While having
features specifically designed for image processing, it can also be viewed as a simplified form of
OpenCL instead of an image processing DSL. With the exception of some restrictions outlined below,
ImageCL programs can contain arbitrary code with a syntax identical to OpenCL C. It can therefore be
used to write general purpose programs, and can work in a standalone fashion without FAST. 

\begin{lstlisting}[caption=Box filter in ImageCL, label=blur, frame=single]
#pragma imcl grid(input)
void blur(Image<float> in, Image<float> out){
    float sum = 0.0;
    for(int i = -1; i < 2; i++){
        for(int j = -1; j < 2; j++){
            sum += in[idx + i][idy + j];
        }
    }
    out[idx][idy] = sum/9.0;
}
\end{lstlisting}

ImageCL is based on the same programming models as OpenCL, but makes two major simplifications, as
well as a number of other changes.

Firstly, in OpenCL, the programmer specifies a two-level thread hierarchy. This concept is
abstracted away in ImageCL, and replaced with a flat thread space. In particular, the user specifies
an \texttt{Image} (described below), and a grid of logical threads, with the same size and
dimensionality as the \texttt{Image} is created. The kernel is the work performed by one such
logical thread, and it is intended to work on its corresponding pixel, although this is not a
requirement.  The built in variables \texttt{idx} and \texttt{idy} store the index of the thread,
and can be used to index the thread-grid defining \texttt{Image} to find the pixel of the thread. If
\texttt{Image}s are not used, the size of the logical thread grid can be specified manually.

Secondly, OpenCL has a complex memory hierarchy, with multiple different logical memory spaces.
These are typically mapped to different hardware memory spaces with different performance
characteristics, as described in Section~\ref{background}. This memory hierarchy is also abstracted
away in ImageCL, the programmer only deals with a single flat address space.

\begin{figure}[h]

    \centering
\begin{subfigure}[htpb]{0.3\textwidth}
    \includegraphics[width=\textwidth]{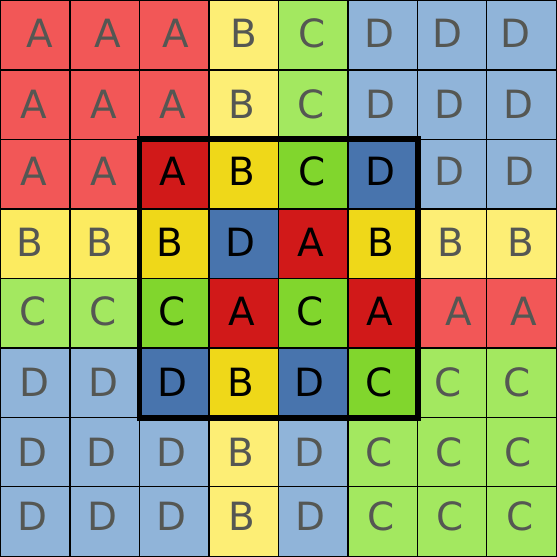}
        \caption{Clamped: values outside are set to that of the closest pixel inside the image. }
\label{border_clamped}
\end{subfigure}
\quad
\begin{subfigure}[htpb]{0.3\textwidth}
    \includegraphics[width=\textwidth]{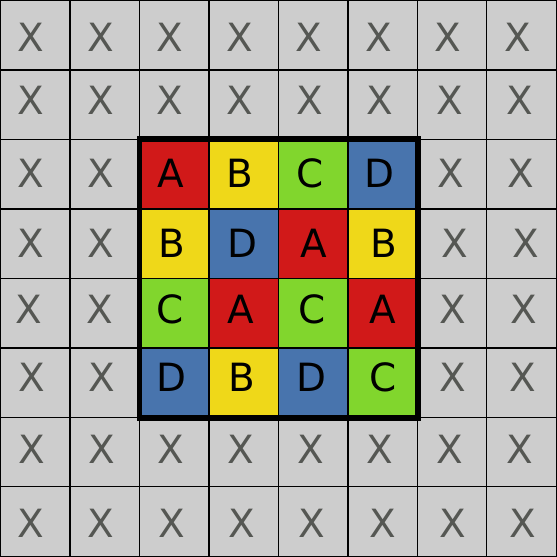}
        \caption{Constant: values outside the image are set to some constant, e.g. 0.}
\label{border_const}
\end{subfigure}
    \caption{Different boundary conditions.}
    \label{border}
\end{figure}

In addition, ImageCL includes the \texttt{Image} data type, intended to
store an image, supports 2D/3D indexing, as shown in Listing~\ref{blur}, and can be
templated with the pixel type. One can also specify different boundary conditions,
making it possible to read outside an \texttt{Image} with well defined results, a situation which
frequently arises in stencil algorithms.  We currently support constant and
clamped boundary conditions, illustrated in Figure~\ref{border}.  \texttt{Image} comes in
addition to, and does not replace, the data types supported by OpenCL, such as regular arrays.

The ImageCL language also includes a small number of compiler directives. The most important is the
already described \texttt{grid} directive, which can be used to determine which \texttt{Image} to
base the thread-grid on, as shown in Listing~\ref{blur}, or the size of the grid directly when no
\texttt{Image}s are used.  Other directives specify boundary conditions, upper bounds
on sizes of arrays, and force optimizations on or off.

The present features of ImageCL are rich enough to express a wide range of parallel image processing
algorithms. However, some features, required for more complex algorithms, are planned for future
versions.  Specifically, there are presently no synchronization or communication primitives (global
barriers can still be achieved by returning control to the host). Furthermore the kernel must be
written as a single function.

\subsection{Implementation}
Our ImageCL source-to-source compiler is implemented using the ROSE \cite{QUINLAN11} compiler
framework. Before being given to the compiler proper, a few typedefs and declarations are added to
the source code (e.g.  declaring \texttt{idx} and \texttt{idy}, declaring an \texttt{Image} class),
necessary to make it valid C++, which ROSE can handle directly. ROSE can generate OpenCL. 

Our compiler can either analyze the input code to find possible tuning parameters, or read values
for the tuning parameters, and apply the relevant transformations to generate OpenCL. The analysis
examines the structure of the abstract syntax tree (AST) and performs various data-flow analyses.
The transformations are applied by modifying the AST.

Since the ImageCL programming model is so close to that of OpenCL, generating naive, unoptimized
OpenCL, is straightforward. It involves replacing \texttt{idx} and \texttt{idy} with thread index
calculations, converting \texttt{Image}s to 1D arrays as well as updating their index calculations,
and adding code to implement the boundary conditions. Finally, OpenCL keywords like
\texttt{\_\_kernel} and \texttt{\_\_global} must be added.

In addition to the kernel code itself, we also generate host code to launch the kernel. We can
either generate host code which can be used as a filter in FAST, or as a standalone function,
callable from any C/C++ application. 

\subsection{Tuning Parameters}

Our tuning parameters are summarized in Table~\ref{params}, and will be described here.

\begin{table}[h]
    \caption{Tuning parameters.}
    \centering
    \begin{tabularx}{0.95\textwidth}{|l|X|}
        \hline
        \textbf{Parameter} & \textbf{Description} \\
        \hline
        Work-group size & The size of a work-group in each dimension. \\\hline
        Thread coarsening & The number of logical threads processed by each real thread
        in each dimension. \\\hline
        Image memory & Whether or not to use image memory. One parameter for each applicable array.  \\ \hline
        Constant memory & Whether or not to use constant memory. One parameter for each applicable array \\ \hline
        Local memory & Whether or not to use local memory. One parameter for each applicable array. \\ \hline
        Thread mapping & To use blocking or interleaved thread mapping. \\ \hline
        Loop unrolling & Loop unroll factor for each applicable loop. \\
        \hline
    \end{tabularx}
    \label{params}
\end{table}

Some of the optimizations, in particular the local memory optimization, are motivated by
computational patterns frequently occurring in image processing, such as stencil computations. They
will therefore be most applicable, and have the largest impact, for such applications.
Furthermore, optimizations involving OpenCL memory hierarchy may have little or no effect on CPUs,
where this hierarchy is placed entirely in main memory. A capable auto-tuner will be
able to handle such effectless parameters.

\subsubsection{Work-Group Size}
As described above, ImageCL has a flat thread space, which must be mapped to the two level thread
hierarchy in OpenCL. The size and shape of the work-groups can have significant impact on
performance \cite{FALCH2}, and are therefore added as tuning parameters. We add a parameter for the
size in each dimension, the dimensionality is the same as the \texttt{Image} upon which the
thread-grid is based.

\subsubsection{Thread Coarsening}
ImageCL has one logical thread for each pixel of the thread-grid defining \texttt{Image}. While good
performance on GPUs require thousands of threads to keep the device busy, using one thread per pixel
can lead to millions of threads, many more than required. It may therefore be beneficial to perform
\emph{thread coarsening} \cite{MAGNI13}, that is, let each thread perform more work, while reducing
the total number of threads. In ImageCL, logical threads can be merged so that each real thread (that is,
each OpenCL work-item) processes a block of pixels. The sizes of this block in each dimension then becomes the tuning
parameters. Not only the amount of work, but also the shape of the block matters, as it
can affect the memory access pattern. Presently, thread coarsening is implemented by wrapping the kernel in for-loops,
thus executing it multiple times. 

\subsubsection{Thread mapping}
When a single real thread (i.e., OpenCL work-item) is used to process multiple logical threads,
as described above, there are several ways to distribute the logical
threads.  Because the logical threads typically work on their corresponding pixel of the thread-grid
defining \texttt{Image}, this mapping can impact memory access
patterns, and thereby, performance.

\begin{figure}[h]
\begin{subfigure}[htpb]{0.3\textwidth}
    \includegraphics[width=\textwidth]{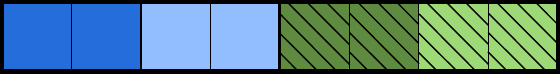}
        \caption{Blocking}
\label{mem_access_blocking}
\end{subfigure}
\hfill
\begin{subfigure}[htpb]{0.3\textwidth}
    \includegraphics[width=\textwidth]{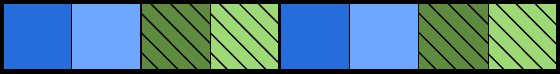}
        \caption{Interleaved}
\label{mem_access_interleaved}
\end{subfigure}

\centering
\begin{subfigure}[htpb]{0.3\textwidth}
    \includegraphics[width=\textwidth]{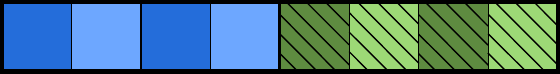}
    \caption{Interleaved in work-group}
\label{mem_access_interleaved_wg}
\end{subfigure}
    \caption{Thread mappings. Cells correspond to logical threads, colors to
    real threads, and pattern to different work-groups.}
\end{figure}

Since the logical threads are organized in an \emph{n}-dimensional grid, a contiguous block of
logical threads can be assigned to each real thread, as illustrated in
Figure~\ref{mem_access_blocking}. While this might give good memory locality, since the pixels of
the logical threads are close, it results in poor \emph{coalescing} on GPUs. Memory
transactions made by different threads on GPUs can be merged, or coalesced, into fewer transactions
if the accesses are close together. It might therefore be better to interleave the logical
threads processed, as shown in Figure~\ref{mem_access_interleaved}. Thus, if the real threads
access the pixels of their logical threads sequentially, they will access a contiguous block of
pixels, leading to coalesced loads. Because of this potential performance impact, we add
whether to use blocked or interleaved thread assignment as a tuning parameter. The implementation of
this parameter simply requires different indexing calculations.

The blocking scheme works well combined with the local memory optimization described below, since
the pixels processed by a work-group still form one contiguous block. This is not the case for the
interleaved scheme. When the local memory optimization is used, the interleaving is therefore
performed within each work-group, as illustrated in Figure~\ref{mem_access_interleaved_wg}.

\subsubsection{Memory Spaces}
In ImageCL, there is only a single address space, by default mapped to OpenCL global memory.  However,
using other OpenCL memory spaces can often affect performance, as described in
Section~\ref{background}. We therefore include whether to place data in other memory spaces as
tuning parameters. The data under consideration are either \texttt{Images},
or general arrays. In the following, we will refer to both as arrays, unless the distinction is
necessary.

\paragraph{Image Memory} can be used in either a read-only or write-only manner. In ImageCL, we
disallow aliasing. We can therefore determine if an array is only read from, or only written to,
by looking at every reference to the array and determine whether it is a read or a write.
To place an array in image memory, we simply change the type declaration, replace the
references with the relevant image memory read or write functions, and change the host memory
allocation.

\paragraph{Constant memory} can only be used in a read-only manner, and has a
limited size. To determine if an array can be placed in constant memory, we therefore check if it is
only read from, as described above, and whether its size is below some threshold. If the size of the
array cannot be determined at compile time, but is known to always be sufficiently small, a compiler
directive can be used to specify this. To place an array in constant memory, we simply add the
relevant address space qualifier.

\paragraph{Local memory} can be used if an \texttt{Image} is only read from and each thread reads from a fixed
size neighbourhood, or stencil, around its pixel, the size of which can be determined at compile
time. Using local memory in this case can increase performance
because the areas read by neighbouring threads will overlap. Loading the data once into local
memory, and then having different threads access it there multiple times may therefore be
beneficial.

To determine the size of the stencil, that is, the area around its central pixel a thread reads
from, we find all the relevant \texttt{Image} references, and make sure they have the form
\texttt{image[idx + c1][idy + c2]}. We then use constant propagation to determine the values of
\texttt{c1} and \texttt{c2}.  Often, \texttt{c1} and \texttt{c2} are
not constants, but depend on the iteration variable of for-loops with a fixed range, as in
Listing~\ref{blur}. In such cases, we use a modified version of constant propagation where we allow
each variable to take on a small set of constant values. If the values of
\texttt{c1} or \texttt{c2} cannot be determined at compile time, the analysis fails, and local
memory is not used. 

\begin{figure}[htbp]
    \centering
    \includegraphics[width=0.55\textwidth]{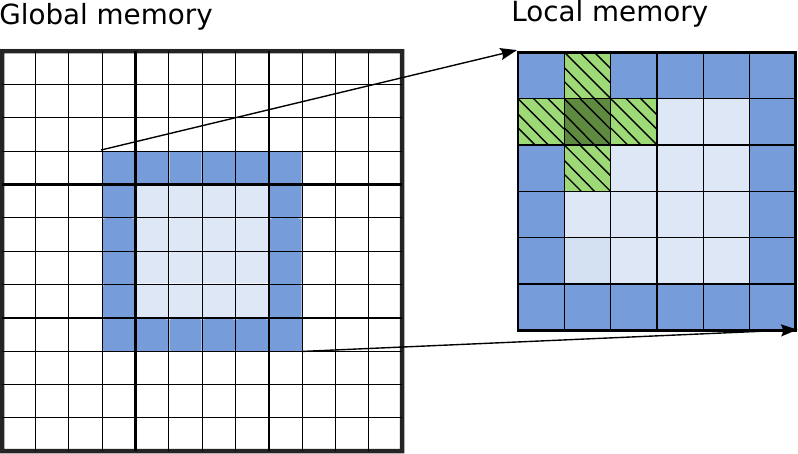}
    \caption{Local memory transformation. The light blue area shows the pixels of a
    work-group, while the green/hatched pixels is the stencil of a single thread. The dark blue area show the additional data
that must be loaded into local memory.}
    \label{local_mem}
\end{figure}

If the local memory optimization is applied, each work-group initially cooperatively loads the
required part of the array into local memory. The required part is the area covered when moving the
center of the bounding box of the stencil over all the pixels belonging to the threads of the
work-group, as illustrated in Figure~\ref{local_mem}. We use the bounding box for simplicity,
although this may cause unnecessary loads. We then replace each load from global memory with a load
from local memory, updating the indexing appropriately.

In general, parts of the stencil might fall outside the \texttt{Image} being read from, and it might
also be the case that the \texttt{Image} read from is smaller than the thread-grid. Since we
restrict this transformation to \texttt{Images}, their boundary conditions ensures that well defined
values can be returned in these cases. Since we only consider cases where \texttt{idx} and \texttt{idy} are
not multiplied, divided, taken the modulo of etc., there is a well defined mapping from the logical
threads of the thread-grid to the pixels of any size array, ensuring that the real threads of a
work-group work on a contiguous area, which can easily be computed. The read-only requirement is
needed to ensure correctness, since we presently do not have any synchronization primitives.

\subsubsection{Loop unrolling}
involves replacing the loop body with multiple copies of itself, while adjusting the
number of iterations accordingly. As this is well known to impact performance, we
add the loop unrolling factor as a tuning parameter.

\section{Results}
\label{results}
To evaluate ImageCL we implemented three image processing
benchmarks, and compared performance with other state of the art solutions. To evaluate
performance portability, we tested on a range of different hardware devices.

\begin{figure*}[t]
    \centering
    \includegraphics[width=\textwidth]{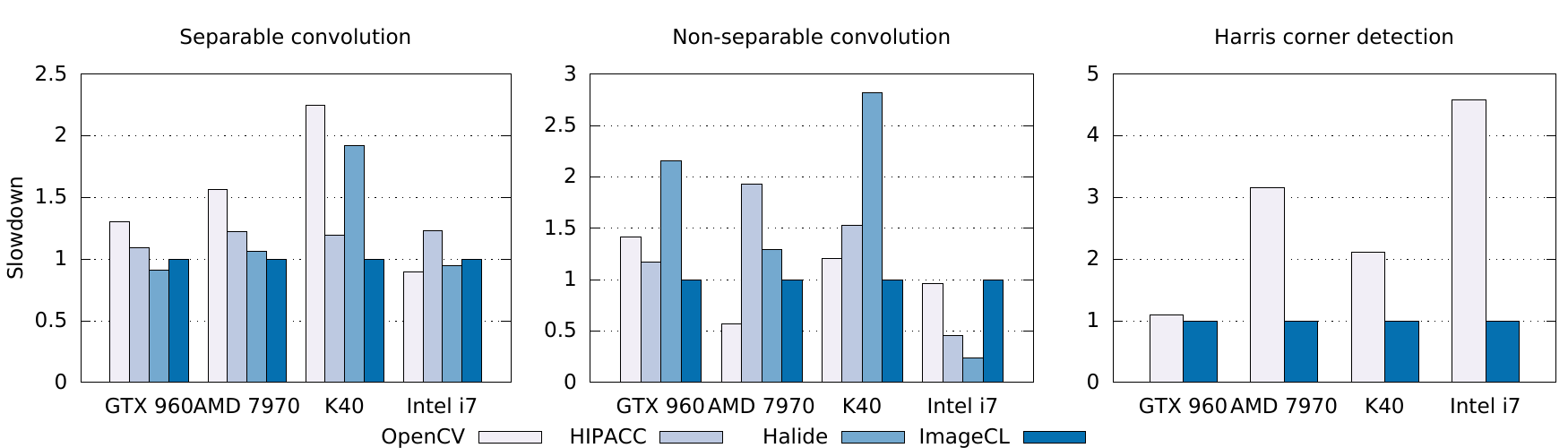}
    \caption{Slowdown compared to ImageCL, computed independently for each benchmark/device
    combination.}
    \label{exectime}
\end{figure*}

The benchmarks are separable convolution, non-separable convolution and Harris corner detection.
Convolution was chosen because it is extensively used in image processing, e.g. for the ubiquitous
Gaussian blurring. It also serves as a proxy for other stencil computations. Harris corner detection
was chosen as an example of a more complex algorithm.

For evaluation, we used three different GPUs, an Nvidia GeForce GTX 960, an Nvidia Tesla K40 and an
AMD Radeon HD 7970 as well as a Intel i7 4771 CPU. 

We compared our performance against Halide \cite{RAGAN2013}, HIPACC \cite{MEMBARTH2016} and OpenCV
\cite{OPENCV}. Halide and HIPACC, described in Section~\ref{background}, are domain specific
languages for image processing, and can generate high performance CPU and GPU implementations.
OpenCV is a widely used library for image processing. It is highly optimized, and contains
implementations for both CPUs and GPUs. We used OpenCV 3.0, the 2015/10/22 release of Halide and
HIPACC 0.8.1.

HIPACC allows the target device and architecture to be specified, so that appropriate optimizations
can be applied. The version of HIPACC we used does not support the AMD 7970. We therefore used the latest
generation of AMD devices it supports instead. While Halide claims that its code can be auto-tuned,
no auto-tuner or auto-tuning interface is distributed with the source code. We therefore performed
extensive manual tuning. As described in Section~\ref{related}, Halide code is divided into a
functional description of the algorithm, and a schedule, describing parallelization, tiling,
vectorization, etc. The manual tuning was therefore carried out by systematically trying out
different possible Halide schedules for each device/benchmark combination.  The ImageCL
implementations were auto-tuned with the machine learning based auto-tuner from our previous work
\cite{FALCH2}, which is described in Section~\ref{compiler}.

Both HIPACC and Halide can generate either OpenCL or CUDA when targeting Nvidia devices. For HIPACC,
we used the CUDA version, and for Halide the OpenCL version, since there performed best. Similarily,
HIPACC can generate OpenCL and C++ code for the CPU. Since the OpenCL versions performed better, we
report those results.

Both Halide and HIPACC can generate specialised code if the values of the filters are known at code
generation time. We evaluated both options, for the separable convolution, the filter values are
known at code generation time, while they are only known at run time for the non-separable convolution.

The execution times reported does not inclue CPU-GPU memory transfers, preparing inputs,
etc., as this will be the same for all alternatives. Due to time constraints, we only compare
against OpenCV for the Harris corner detection.

Figure~\ref{exectime} shows the results.
For separable convolution, we used a 4096x4096 image with pixels of type \texttt{float}, a 5x5
filter, and constant boundary condition. ImageCL was faster than the alternatives on the GPUs,
achieving speedups between 1.06 and 2.25, with the sole exception of Halide on the GTX 960, which
was 9.1\% faster than ImageCL. On the CPU ImageCL was 1.05 and 1.11 times slower than Halide and
OpenCV, but was 1.23 times faster than HIPACC.

For non-separable convolution, we used a 8192x8192 image with pixels of type
\texttt{unsigned char}, a 5x5 filter, and clamped boundary condition. ImageCL was faster than the
alternatives on the GPUs, achieving speedups between 1.17 and 2.82, with the sole exception of
OpenCV on the AMD 7970, which was 43.4\% faster than ImageCL. On the CPU, ImageCL performed worst,
only 1.06 times slower than OpenCV, but 4.24 times slower than Halide.

For Harris corner detection, we used a 5120x5120 image with pixels of type \texttt{float}, and a
block size of 2x2. ImageCL significantly outperformed OpenCV on the AMD 7970, K40 and Intel i7,
achieving speedups of 3.15, 2.11 and 4.57 respectively. On the GTX 960, the performance was more
similar, with ImageCL achieving a speedup of 1.08.

Table~\ref{bestconfigs} shows the parameters, and the values found, by the auto-tuner for
separable convolution.
Table~\ref{bestconfigs_nonsep}, \ref{bestconfigs_sobel} and \ref{bestconfigs_harris} shows the same data for
the non-separable convolution, and the Sobel and Harris kernels of the Harris corner detection, respectively.

\begin{table}[h]
\centering
\caption{Configurations found by auto-tuner for the row (R) and column (C) kernels of the separable convolution.}
\label{bestconfigs}
\begin{tabular}{|l|l|l|l|l|l|l|l|l|}
\hline
\textbf{Device} &\multicolumn{2}{l|}{\textbf{AMD 7970}} & \multicolumn{2}{l|}{\textbf{GTX 960}} & \multicolumn{2}{l|}{\textbf{K40}} & \multicolumn{2}{l|}{\textbf{Intel i7}} \\ \hline
\textbf{Kernel}              &  R   &  C    & R     & C    & R    & C     & R    &  C   \\ \hline
Px./thread X         &  4   &  2    &  1    & 1    & 2    &   2   & 128  &  32  \\ \hline
Px./thread Y         &  1   &  2    &  1    & 2    & 1    &   2   & 1    &  1  \\ \hline
Work-group X   &  64  &  16   &  16   & 64   & 16   &   16  & 8    &  16  \\ \hline
Work-group Y   &  4   &  16   &  16   & 4    & 16   &   16  & 1    &  2  \\ \hline
Interleaved         &  1   &  1    &  0    & 0    & 0    &   0   & 1    &  1  \\ \hline
Image mem.        &  0   &  1    &  1    & 0    & 1    &   1   & 0    &  0  \\ \hline
Local mem.        &  1   &  1    &  0    & 0    & 0    &   0   & 0    &  0  \\ \hline
Constant mem.     &  1   &  1    &  1    & 1    & 1    &   1   & 1    &  1  \\ \hline
Unroll loop 1       &  0   &  1    &  0    & 0    & 1    &   0   & 1    &  0  \\ \hline
\end{tabular}
\end{table}

\begin{table}[h]
\centering
\caption{Configurations found by auto-tuner for the non-separable convolution.}
\label{bestconfigs_nonsep}
\begin{tabular}{|l|l|l|l|l|}
\hline
\textbf{Device} & \textbf{AMD 7970} & \textbf{GTX 960} & \textbf{K40} & \textbf{Intel i7} \\ \hline
Px/thread X         &  4   &  4    &  4    & 256     \\ \hline
Px/thread Y         &  16   &  4    &  8    & 2     \\ \hline
Work-group X   &  64  &  8   &  32   & 2    \\ \hline
Work-group Y   &  4   &  32   &  4   & 8     \\ \hline
Interleaved         &  0   &  1    &  0    & 1     \\ \hline
Image mem        &  0   &  0    &  1    & 0     \\ \hline
Local mem        &  0  &  1    &  0    & 0     \\ \hline
Constant mem     &  1   &  1    &  1    & 1     \\ \hline
Unroll loop 1       &  1   &  1    &  1    & 1      \\ \hline
Unroll loop 2       &  0   &  1    &  1    & 1      \\ \hline
\end{tabular}
\end{table}

\begin{table}[h]
\centering
\caption{Configurations found by auto-tuner for the Sobel kernel of the Harris corner detection.}
\label{bestconfigs_sobel}
\begin{tabular}{|l|l|l|l|l|}
\hline
\textbf{Device} & \textbf{AMD 7970} & \textbf{GTX 960} & \textbf{K40} & \textbf{Intel i7} \\ \hline
Px/thread X         &  1   &  4    &  1    & 32     \\ \hline
Px/thread Y         &  1   &  2    &  4    & 4     \\ \hline
Work-group X   &  128  &  32   &  32   & 64    \\ \hline
Work-group Y   &  1   &  2   &  4   & 1     \\ \hline
Interleaved         &  0   &  1    &  0    & 0     \\ \hline
Image mem        &  0   &  0    &  1    & 0     \\ \hline
Local mem        &  0  &  1    &  0    & 0     \\ \hline
\end{tabular}
\end{table}

\begin{table}[h]
\centering
\caption{Configurations found by auto-tuner for the Harris kernel of the Harris corner detection.}
\label{bestconfigs_harris}
\begin{tabular}{|l|l|l|l|l|}
\hline
\textbf{Device} & \textbf{AMD 7970} & \textbf{GTX 960} & \textbf{K40} & \textbf{Intel i7} \\ \hline
Px/thread X        &  1   &  1    &  32    & 8     \\ \hline
Px/thread Y        &  1   &  2    &  2     & 2     \\ \hline
Work-group X       &  32  &  128   &  64   & 32    \\ \hline
Work-group Y       &  8   &  2   &  16   & 2     \\ \hline
Interleaved        &  0   &  0    &  1    & 0     \\ \hline
Image mem dx       &  1   &  1    &  1    & 0     \\ \hline
Image mem dy       &  1   &  1    &  1    & 0     \\ \hline
Local mem dx       &  1  &  1    &  0    & 0     \\ \hline
Local mem dy       &  1  &  1    &  0    & 0     \\ \hline
Loop 1             &  1  &  0    &  0    & 1     \\ \hline
Loop 2             &  0  &  1    &  0    & 0     \\ \hline
\end{tabular}
\end{table}

\section{Discussion}
\label{discussion}
ImageCL performs comparatively well because it is able to apply a wide range of optimizations, and
use auto-tuning to pick the correct optimizations for a given device and benchmark. For example, the
good performance compared to Halide on the K40 is caused in part by ImageCL using image
memory, an optimization Halide does not expose and therefore cannot be applied even if the
programmer suspects it might help. The varying and sometimes poor performance of OpenCV illustrates
the problem of performance portability. It has separate implementations for the CPU and GPUs, a
solution that requires extra work and scales poorly. For the GPUs, it is increasingly difficult
to write a single implementation that performs well on all of them. ImageCL's use of auto-tuning
represents a more robust and scalable solution.

However, some optimizations cannot be applied in the current version of ImageCL.
Halide achieves good performance, in particular on the GTX 690 by merging the two 
separable convolution kernels, caching the intermediary result in local memory. Due to the
lack of synchronization and communication primitives, this optimization can currently not be applied
in ImageCL. We suspect that the poor results for non-separable convolution on the CPU is caused
by lack of vectorization. Our compiler does currently not perform vectorization, instead relying on the
vectorization capabilities of the OpenCL runtime, which performed better for the
separable convolution and corner detection, than the non-separable convolution.

The poor performance for non-separable convolution on the CPU is also caused by the implementation of the
clamped boundary condition. If the constant boundary condition is used, the execution time is
reduced by a factor of 2. As such, it is not a fundamental flaw of our approach, and
using auto-tuning to pick different implementations of the boundary condition for different devices
could overcome this issue.

Despite good results, the time required for auto-tuning, even
using our machine learning based auto-tuner \cite{FALCH2}, can be a significant drawback. For the
casses presented, the auto-tuner executed around 1700 valid candidate implementations during its search
for each device/benchmark combination. This required around 2 hours
in total. For each candidate implementation, running the actual kernel takes milliseconds,
additional time is required to prepare inputs and transfer data to and from the GPU. More
significantly, each candidate implementation must be compiled by our compiler and the OpenCL
compiler which can take up to 1-3 seconds. In comparison, OpenCV and HIPACC have no tuning overhead,
while the manual tuning for Halide required several hours. Because of the overhead, this kind of
auto-tuning will be most usefull for code that is executed many times, like library functions. The
good results achieved here, as well as in our previous work \cite{FALCH2} gives us
some confidence in our auto-tuners abilities, but it does not give any guarantees about how close the
solution found is to the globally best one, which can only be found by time consuming exhaustive
search. Auto-tuning might be more challenging for larger applications with
more complex search spaces.

\section{Conclusion and Future Work}
\label{conclusion}
The increasing popularity of heterogeneous computing has made performance portability, making code
execute with good performance when moved between different devices, harder to achieve. In this
paper, we have introduced the ImageCL language and its source-to-source compiler. ImageCL is a simplified form of OpenCL with features for image processing, and is translated to optimized OpenCL. Our source-to-source
compiler can apply various optimizations to generate multiple candidate
implementations. An auto-tuner can then pick the best implementation for a given device.
We have evaluated the performance of ImageCL with three image processing benchmarks, and are able
outperform state of the art solutions in several cases, achieving speedups of up to 4.57x.  Our
proposed solution therefore shows promise as a way to come closer to the goal of heterogeneous
performance portability.

Future work may include adding additional features, such as synchronization and communication
primitives, and increased support for vectorization. Another possibility is
developing a multi-device or distributed version using MPI. Finally, we intend to evaluate ImageCL
with more diverse benchmarks, on a broader range of hardware, including CPUs and accelerators.

\section{Acknowledgements}
The authors would like to thank NTNU IME and NTNU MedTech for their support of the Medical
Computing \& Visualization project, Nvidia's GPU Research Center program and NTNU for hardware
donations to our NTNU/IDI HPC-Lab, and Drs. Malik M. Z. Khan and Erik Smistad for their helpful
discussions. The authors would also like to thank Dr. Keshav Pingali and ICES at the University of Texas at
Austin for hosting us during the work for this article.

\bibliographystyle{plain}
\bibliography{referanser}

\end{document}